\documentclass[aps,prl,reprint,superscriptaddress]{revtex4-2}

\usepackage{amsmath, amsthm, amsfonts, amssymb,amstext}
\usepackage{mathptmx}
\usepackage{bm}
\usepackage{xfrac}
\usepackage{mathrsfs}
\usepackage{latexsym}
\usepackage{array}

\usepackage{soul}

\usepackage{graphicx}

\usepackage{amsmath,amssymb,amsfonts}

\usepackage{enumitem}

\usepackage{xcolor}

\usepackage{soul,xcolor}

\usepackage{tikz,tkz-euclide}

\usepackage[colorlinks=true,citecolor=blue,linkcolor=blue,urlcolor=blue]{hyperref}

\newcommand{\abs}[1]{\left|#1\right|}

\let\ni\noindent
\usepackage{physics}
\usepackage{verbatim}
\usepackage{array}
\usepackage{tabularx}

\usepackage{tikz,ifthen} 
\usetikzlibrary{shapes,arrows,positioning,automata,backgrounds,calc,er,patterns,matrix}

\begin{document}
\begin{abstract}

Recent advances in the field of quantum technologies have opened up the road for the realization of small-scale quantum simulators of lattice gauge theories which, among other goals,  aim at improving our understanding on the non-perturbative mechanisms underlying the confinement of quarks. In this work, considering periodically-driven arrays of Rydberg atoms  in a tweezer  ladder geometry, we devise a scalable  Floquet scheme for the  quantum simulation  of the real-time dynamics in a  $\mathbb{Z}_2$ LGT. Resorting to an external magnetic field to tune the angular dependence of the Rydberg dipolar interactions, and by a suitable tuning of the driving parameters, we manage to suppress the main gauge-violating terms, and show that an  observation of gauge-invariant confinement dynamics in the Floquet-Rydberg setup is at reach of current experimental techniques. Depending on the lattice size, we present a thorough numerical test of the validity of this scheme using either exact diagonalization or matrix-product-state algorithms for the periodically-modulated real-time dynamics.
\end{abstract}

\title{A Floquet-Rydberg quantum simulator for confinement in  $\mathbb{Z}_2$ gauge theories}

\author{Enrico C. Domanti}

\email{Enrico.Domanti@tii.ae}
\affiliation{Quantum Research Centre, Technology Innovation Institute, Abu Dhabi, UAE}

\affiliation{Dipartimento di Fisica e Astronomia, Via S. Sofia 64, 95127 Catania, Italy}

\affiliation{INFN-Sezione di Catania, Via S. Sofia 64, 95127 Catania, Italy}

\author{Dario Zappalà}
\affiliation{INFN-Sezione di Catania, Via S. Sofia 64, 95127 Catania, Italy}
\affiliation{Centro Siciliano di Fisica Nucleare e Struttura della Materia, Via S. Sofia 64, 95127 Catania, Italy}

\author {Alejandro Bermudez}
\altaffiliation[On  sabbatical at ]{Department of Physics, University of Oxford, Clarendon Laboratory, Parks Road, Oxford, UK.}

\affiliation{Instituto de Física Teorica, UAM-CSIC, Universidad Autonoma de Madrid, Cantoblanco, 28049 Madrid, Spain}

\author {Luigi Amico}

\affiliation{Quantum Research Centre, Technology Innovation Institute, Abu Dhabi, UAE}

\affiliation{Dipartimento di Fisica e Astronomia, Via S. Sofia 64, 95127 Catania, Italy}

\affiliation{INFN-Sezione di Catania, Via S. Sofia 64, 95127 Catania, Italy}

\affiliation{Centre for Quantum Technologies, National University of Singapore, 3 Science Drive 2, Singapore 117543, Singapore}

\maketitle

\date{\today}

{\textit{Introduction}.--}
Gauge theory provides us with the basic language to understand the fundamental interactions~\cite{Peskin:1995ev}. Such theories arise from promoting global symmetries, e.g. $SU(3)$ ($SU(2)_L \times U(1)$) for the strong (electroweak) interactions, to local ones by the introduction of gauge fields coupled to matter ~\cite{PhysRev.96.191}. The discretization  of gauge theories  on a space-time lattice not only provides  a natural  cutoff, but a means to go  beyond perturbative calculations~\cite{PhysRevD.10.2445}, which is crucial to understand 
 phenomena such as quark confinement~\cite{greensite_2020}. In spite of the enormous progress of imaginary-time Monte Carlo  methods for lattice gauge theories (LGTs), epitomised by the    verification of the quark-model prediction of the hadron masses~\cite{doi:10.1126/science.1163233}, finite-density and real-time phenomena  are still beyond reach
 ~\cite{NAGATA2022103991}. An approach to overcome these limitations arises from realising that the lattice  is not compelled to be a mathematical construct but, may indeed have  a physical reality. Recent advances in quantum technology provide an effective avenue to implement the above program. By a careful  experimental design, one can tailor the lattice and control the real-time matter dynamics to mimic the target model, thus realizing a LGT quantum simulator (QS)~\cite{Feynman_1982,Cirac2012}. These QSs can be thought of as special-purpose quantum computers~\cite{nielsen00}, and their use for LGTs has recently raised the interest of a broad and  diverse community (see the recent reviews~\cite{https://doi.org/10.1002/andp.201300104,Zohar_2016,doi:10.1080/00107514.2016.1151199,Bañuls2020,Carmen_Bañuls_2020,doi:10.1098/rsta.2021.0064,Klco_2022, https://doi.org/10.48550/arxiv.2204.03381}).

\begin{figure}[th!]
    \centering
    \includegraphics[width = \linewidth]{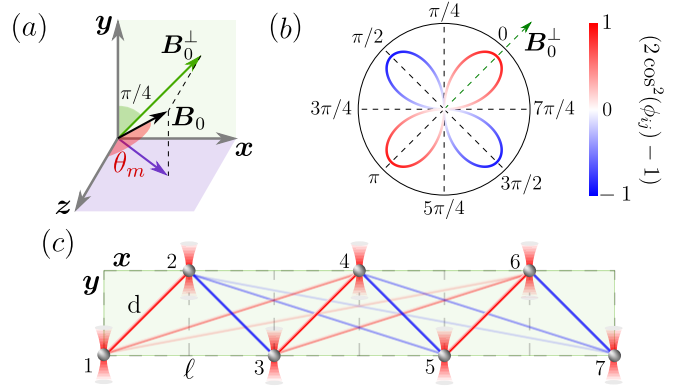}
    \caption{\textbf{
Rydberg tweezer array:}
(a) The external magnetic field $\boldsymbol{B}_0$ makes an  angle $\theta_m = 54.7^{\rm o}$ 
with respect to the $\boldsymbol{z}$ axis. The projection of $\boldsymbol{B}_0$ onto the $xy$ plane where the atoms reside, $\boldsymbol{B}_0^\perp$, makes an angle of $45^{\rm o}$ with the $\boldsymbol{x}$ axis. (b) Angular distribution of the XY couplings as a function of the angle  $\phi_{ij}$ between the interatomic vector $\boldsymbol{R}_{ij}$ and $\boldsymbol{B}_0^\perp$, which vanish at the critical angles $\phi_{ij} = \pm 45^{\rm o}, \pm \,135^{\rm o}$. (c) Ladder configuration of the Rydberg atoms trapped in optical tweezers. The atoms are arranged on the vertices of isosceles right triangles of sides $d$ and base $b = \sqrt{2} \, d$. The interactions are represented as colored lines according to the color-scheme of panel (b). Dashed lines highlight the critical directions of the vanishing XY terms, which forbid a direct coupling between even-even and odd-odd spins. In this paper, we drive the odd atoms detuning according to $H_\text{drive} = \sum_{i \, \text{odd}} \frac{\eta \omega_d}{2} \cos(\omega_d \, t + \varphi_i) \sigma_i^z$, with $\eta \approx 2.4$ and $\phi_{2i+1} = i \frac{\pi}{2}$.}
    \label{fig:lattice}
\end{figure}

In spite of  recent experimental progress~\cite{https://doi.org/10.1002/andp.201300104,Zohar_2016,doi:10.1080/00107514.2016.1151199,Bañuls2020,Carmen_Bañuls_2020,doi:10.1098/rsta.2021.0064,Klco_2022, https://doi.org/10.48550/arxiv.2204.03381} current technologies are still far from allowing
large-scale QSs of the standard model of particle physics. This limitation is either due to the accumulation of errors for QSs operating in  digital mode, or to the limited flexibility of QSs operating in the analog mode. To exploit the full potential of analog QSs, which are in principle more amenable for scaling in the presence of errors~\cite{Flannigan_2022,trivedi2022quantum},  it is of primary importance to devise novel schemes that engineer the high-weight  terms characteristic of gauge theories in  timescales that are faster than current decoherence sources.  To devise these fundamental schemes as building blocks of more complicated models, the community is focusing on  models in reduced spacetime dimensions and simpler gauge groups~\cite{PhysRev.128.2425, COLEMAN1975267,THOOFT1974461, Gross_1993,wegnerz2}, which can still provide important insights. The $\mathbb{Z}_2$ LGT is a paradigmatic case in this regard~\cite{wegnerz2}. On one hand, it is a playground to    understand  confinement (deconfinement) in (1+1) dimensional chains~\cite{PhysRevLett.124.120503,PhysRevLett.127.167203} (ladders~\cite{PhysRevX.10.041007,PhysRevResearch.3.013133}). At the same time,  the $\mathbb{Z}_2$ LGT in 2+1 dimensions  serves to understand central questions in condensed matter as topological order~\cite{KITAEV20032, Wen_2013}, confined and Higgs phases~\cite{PhysRevD.19.3682,PhysRevB.82.085114}, and the interplay of superconductivity and charge deconfinement~\cite{Gazit2017,doi:10.1073/pnas.1806338115}. In the context of QSs, targeting these simpler LGTs dispenses with additional complications in higher dimensions and non-Abelian gauge groups.  Although  LGTs can sometimes be simulated effectively, e.g. integrating the gauge fields~\cite{Muschik_2017, PhysRevResearch.2.023015}  or vice versa~\cite{https://doi.org/10.48550/arxiv.2206.00685,https://doi.org/10.48550/arxiv.2206.08909}, developing schemes  where matter and gauge fields correspond to physical degrees of freedom of the QS  is  essential  to observe their interplay in confinement, as emphasised in recent proposals based on neutral atoms~\cite{Barbieroeaav7444} and trapped ions~\cite{bazavan2023synthetic}. Although these proposals can be in principle scaled up~\cite{PhysRevLett.121.223201,PhysRevResearch.2.013288,PhysRevX.10.021057,PhysRevLett.129.160501,Ohler_2022, mildenberger2022probing,Homeier2023,doi:10.1073/pnas.2304294120}, so far, experiments have demonstrated only building blocks of the full theory~\cite{z2LGT_2,gan_2020}. 

In this work, we provide a scalable  analog QS for the $\mathbb{Z}_2$ LGT based on Rydberg atoms in optical tweezers~\cite{Browaeys_2016,10.1116/5.0036562}. We shall focus on a scheme which encodes a spin-$1/2$ variable  in a pair of nearby  Rydberg levels of opposite  parity $\ket{\uparrow}_i=\ket{r}_i$,  $\ket{\downarrow}_i=\ket{r'}_i$. In this case, the dipolar interactions are described by an XY model with long-range ${1}/{R^3}$ couplings. Below, we  describe how to obtain the desired LGT   by  periodically driving the  atoms, exploiting Floquet engineering~\cite{RevModPhys.89.011004} in a way that effectively gauges the  global symmetry of the XY model, 
transforming it to  a local $\mathbb{Z}_2$  gauge symmetry, where some atoms represent matter, and some others   gauge fields.

{\it Target model}.-- Our goal is to achieve the  $\mathbb{Z}_2$ LGT %
\begin{equation}
    \label{eq:z2model}
    {H}_{\mathbb{Z}_2}\!\! =\!\! \sum_{n=1}^N   \!J_{\rm t}\!\left(\!\!a_n^\dagger \, \tau^z_{n+\frac{1}{2}}a_{n+1}^{\phantom{\dagger}} + {\rm H.c.}\!\!\right)\!\!  +\mu(-1)^na_n^\dagger a_{n}^{\phantom{\dagger}}+ h   \tau^x_{n+\frac{1}{2}} ,
\end{equation}
where $J_{\rm t}$ is the tunneling strength, $h$ ($\mu$) plays the role of the   electric-field coupling strength (particle mass). The matter content corresponds to   hardcore bosons that can be created (annihilated)  $a_{n}^\dagger (a_{n}^{\phantom{\dagger}})$ at the lattice sites, such that double occupancies are forbidden $(a_{n}^\dagger)^2= (a_{n}^{\phantom{\dagger}})^2=0$. The  gauge fields are described by magnetic-type (electric-type)  Pauli operators $\tau^z_{n+{1}/{2}} (\tau^x_{n+{1}/{2}})$  on the links, allowing for a local symmetry $\displaystyle{[H_{\mathbb{Z}_2},G_n] = 0,\hspace{2ex} G^{\phantom{\dagger}}_n = \tau_{n-{1}/{2}}^x \, {\rm exp}\{{\rm i}\pi a_n^\dagger a_n^{\phantom{\dagger}}\} \, \tau_{n+{1}/{2}}^x}$. As customary, hardcore  bosons   can be mapped to fermions through a Jordan-Wigner transformation $a_n=\Pi_{m<n}{\rm e}^{{\rm i}\pi c_m^\dagger c_m}c_n$, such that a further phase transformation $c_n\rightarrow c_n {\rm e}^{{\rm i}\frac{\pi}{2}n}$   maps Eq.~\eqref{eq:z2model} onto the  Kogut-Susskind $\mathbb{Z}_2$ LGT on a chain~\cite{PhysRevD.11.395,PhysRevD.16.3031}.

{\it Nearest-neighbor Floquet scheme.--} Before delving into the additional complexity of the dipolar Rydberg case, we illustrate our Floquet scheme for nearest-neighbor interactions. This case corresponds to the XY chain~\cite{LIEB1961407} with  dynamical  longitudinal $h_l^ i$ and transverse $h_t^i$ fields%
\begin{equation}
\label{eq:model}
    {H}_{\rm XY} =  \sum_{i=1}^{N_a} \!\!J \bigg(  \sigma_i^+ \sigma_{i+1}^- + {\rm H.c.}\bigg)+h_l^i\sigma_i^z +h_t^i \sigma_i^x ,
\end{equation}
where $\sigma^\alpha_i\,, \alpha=x,z,\pm $ are diagonal and ladder operators of the Pauli algebra, and we take $\hbar=1$ such that all couplings in \eqref{eq:model} have units of Hz. Since we aim at generating 3-body terms~\eqref{eq:z2model} from 2-body interactions~\eqref{eq:model}, second-order processes $\mathcal{O}(J^2)$ should outweigh first-order ones $\mathcal{O}(J)$. Moreover,  to achieve gauge invariance, the scheme should single out a specific set of  terms from all  possible  second-order processes. We shall see that both of the above problems can be overcome through a selective driving protocol. 

Specifically we consider a Floquet scheme that drives the odd spins with a time-periodic longitudinal field   of frequency $\omega_{\rm d}$, strength $\eta\omega_{\rm d}$, and phase $\varphi_i$, together with a   static staggered part $h_l^i=\frac{1}{2}[(-1)^i\delta h+\eta\omega_{\rm d}\cos(\omega_{\rm d}t+\varphi_i)]\delta_{i,\text{odd}}$. The even spins  are instead subject to a static transverse field $h_t^i=\frac{\Omega}{2}\delta_{i,\text{even}}$. To achieve the aforementioned selective dressing, we first need to  inhibit  the gauge-breaking first-order terms. For this, we exploit a spin-version of the  coherent destruction of tunneling~\cite{PhysRevB.34.3625,PhysRevLett.99.220403}, through which $\mathcal{O}(J)$ terms get dressed by the absorption/emission of energy packets  $\epsilon_\ell=\ell\omega_{\rm d}$, $\ell\in\mathbb{Z}$,  from/into the driving field. By using a very fast modulation $\omega_{\rm d}\gg J$, these processes become off-resonant,   and the  XY coupling gets dressed by the $\ell=0$ term as $J\to J\mathsf{J}_0(\eta)$. Thus, by setting the relative amplitude to a zero of this Bessel function, the first-order terms get suppressed.

\begin{figure}
    \centering
    \includegraphics[width = 0.98 \linewidth]{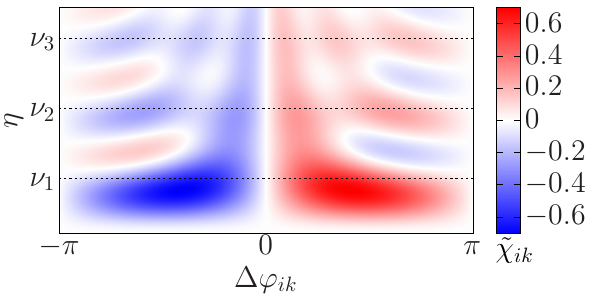}
    \caption{\textbf{  Dressing parameter:}  $\tilde{\chi}_{ik} = -\rm i \, \chi_{ik}$ is depicted as a function of the phase difference $\Delta\varphi_{ik}$ and relative driving strength $\eta$. Here $\nu_n$ denotes the $n$-th zero of $\mathsf{J}_0(x)$. For $\eta = \nu_1$, $\tilde{\chi}_{ik}$ has a maximum (minimum) at $\Delta\varphi_{ik} = 60^{\rm o}$ ($-60^{\rm o}$).}
    \label{fig:chi}
\end{figure}

At second order $\mathcal{O}(J^2/\omega_{\rm d})$, we get three types of processes: assisted tunnelings between  odd sites $\sigma_{2n-1}^+\sigma^z_{2n}\sigma_{2n+1}^-$, assisted tunnelings between  even sites $\sigma_{2n}^+\sigma^z_{2n+1}\sigma_{2n+2}^-$, and two-body terms arising from a back and forth spin flips. 
The effective Hamiltonian reads
\begin{equation}
\label{eq:nn_eff_h}
    H_{\rm eff}=\!\sum_{i \text{ even}}\!\! \frac{\Omega}{2}\sigma_i^x +\!\sum_{i \text{ odd}}\!\! (-1)^i \frac{\delta h}{2}\hspace{0.2ex}\sigma_{i}^z+\!\!\sum_{\langle i,j,k\rangle}\!\!\frac{J^2}{\omega_{\rm d}}\chi_{ik}\sigma_i^+\sigma_{j}^z\sigma_k^-,
\end{equation}
where $\langle i,j,k\rangle$ is a  nearest-neighbour triplet. Here,  we have introduced a  dressing parameter that contains all  second-order processes where $\ell$ 'quanta' are virtually absorbed from and  emitted into the driving field
\begin{equation}
\label{eq:chi}
\chi_{ik}=\sum_{\ell>0}\frac{2{\rm i}}{\ell}\mathsf{J}^2_\ell(\eta)\sin[\ell(\varphi_k-\varphi_i)].
\end{equation}
We note that, in spite of the sum in Eq.~\eqref{eq:nn_eff_h} including all possible triplets, $\chi_{ik}$ is non-zero only for $i\neq k$ both odd, as all other cases 'see' the same driving phase $\varphi_i-\varphi_k=0\,{\rm mod}2\pi$. We thus achieve the desired selectivity by a destructive interference, cancelling all second-order  processes but those with $\varphi_i\neq\varphi_k$. These processes correspond to the tunneling of an excitation  between odd spins depending on the population of the intermediate even spin (see  Supplemental Material). In order to maximise such assisted tunneling, it is convenient to choose $\eta = \nu_1 \approx2.405$ and $\varphi_{2n+1} = n \frac{\pi}{3}$, where $\nu_1$ denotes the first zero of the Bessel function $\mathsf{J}_0(\eta)$ (see Fig.~\ref{fig:chi}). With this choice $\chi_{2n-1,2n+1} = \chi \approx - 0.63 \, \rm i$, the dressed second-order dynamics is slower by a factor of $|J\chi/\omega_{\rm d}|\approx 0.06$. Eq.~\eqref{eq:z2model} can be finally obtained by identifying matter (gauge fields) with the odd (even) spins, such that $N_a=2N+1$: $a_n^{\phantom{\dagger}}=\sigma_{2n+1}^-,\hspace{0.5ex}a_n^{{\dagger}}=\sigma_{2n+1}^+,\hspace{0.5ex}
\tau^z_{n+1/2}=\sigma_{2n}^z,\hspace{0.5ex} \tau^x_{n+1/2}=\sigma_{2n}^x$,
together with the following   parameters 
\begin{equation}
\label{eq:micro_param}
J_{\rm t}=\frac{J^2}{\omega_{\rm d}}\chi, \hspace{2ex} h=\frac{\Omega}{2},\hspace{2ex}\mu=\delta h.
\end{equation}

\begin{figure}
    \centering
    \includegraphics[width = 1.0\linewidth, angle=0]{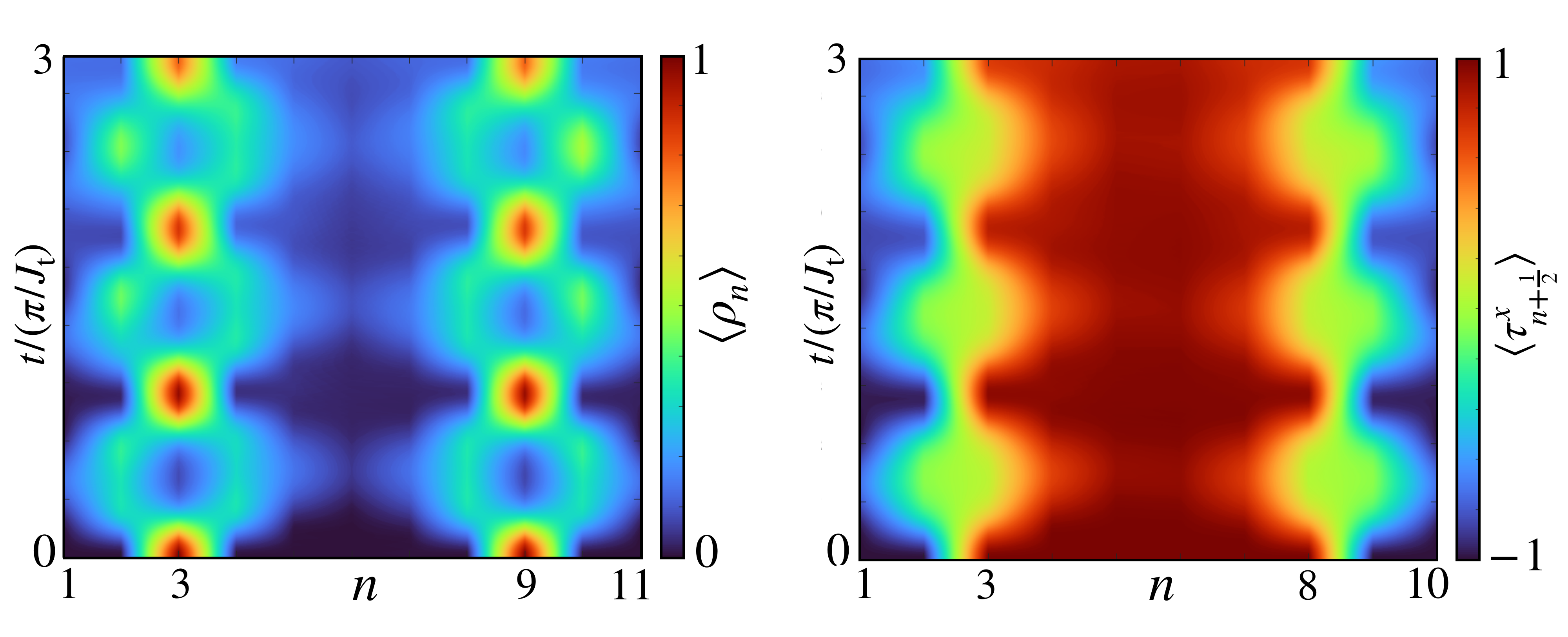}
    \caption{\textbf{ $\mathbb{Z}_2$ confinement:} Two particles are initialized at positions $i=3,9$ and connected by an electric-field string. Here, $h = J_{\rm t}$ and $\mu = 0$. Left and right panels correspond to the contour plots of the charge distribution  $\langle \rho_n(t)\rangle=\bra{\psi(t)} a_n^\dagger a_n\ket{\psi(t)}$ and the electric field $\langle\tau_{n+1/2}^x(t)\rangle$, where the evolution of $\ket{\psi(t)}$ under ${H}_{\rm XY}(t)$ with $\omega_d = 30J$ is approximated using MPS with bond dimension $= 20$.}
    \label{fig:oscillations}
\end{figure}

After presenting our Floquet QS scheme, we provide numerical benchmarks that support its validity. We consider an initial state that allows to discuss confinement. In the LGT language, such a state corresponds to a pair of dynamical charges  at positions $n_1$ and $n_2$  connected by an electric-field string $\ket{\psi_0}=\ket{0,- \dots -,1_{n_1}, +, 0,+\dots  +,1_{n_2},-,0,- \dots -,0}$, where $\tau^x\ket{\pm}=\pm \ket{\pm}$. This initial state belongs to the neutral gauge sector $G_n\ket{\psi_0} = \ket{\psi_0} \, \forall n\neq 1, N$. We consider the case $\Omega \neq 0$, $\delta h = 0$, and simulate the real-time dynamics of the periodically-modulated XY chain~\eqref{eq:model} of $2N+1=21$ spins using matrix-product-state algorithms (TEDB)\cite{itensor,itensor-r0.3,perez2006matrix,banuls2023tensor}. In Fig.~\ref{fig:oscillations}, we display the obtained time-evolution: indeed the charges do not spread indefinitely but instead perform periodic oscillations around their initial positions, 
accompanied  by the  stretching and compressing of the electric string. Indeed, when $\Omega \neq 0$, the effective $\mathbb{Z}_2$ model~\eqref{eq:z2model} is characterized by a non-zero electric coupling $h = \frac{\Omega}{2}$. In this case, pairs of particles separated by a positive electric string experience an attracting potential that grows linearly with the string length, and are confined into meson-like bound states \cite{Surace_2021,borla2020confined}. The relative coordinate wave-function solve a Wannier-Stark equation and thus Bloch oscillations arise \cite{Surace_2021,domanti2023coherence,bazavan2023synthetic}. The observed dynamics is effectively restricted to the neutral gauge sector, in agreement with Eq.\eqref{eq:z2model}, within errors which come from higher-order Floquet terms. The amount of gauge-breaking grows with time, but remains below $15 \%$ violation of the condition $G_n\ket{\psi_0} = \ket{\psi_0} \, \forall n\neq 1, N$ - see Supplemental Material. 

{\it Dipolar  Floquet-Rydberg  scheme.--} After presenting  the ideal nearest-neighbor Floquet  scheme, we provide details for the implementation through Rydberg-atom arrays~\cite{Browaeys_2016,10.1116/5.0036562}. 
We note that the Rydberg levels $\ket{r}$ and $\ket{r'}$ we refer to are $r=nL_{J},m$ and $r'=n'L'_{J'},m'$, having opposite parity and  dipole-allowed transitions, e.g. $ L=S {\,\,\rm or \,\,}P$   and $ L'=P {\,\,\rm or \,\,}D$ with $\Delta J,\Delta m\in\{0,\pm 1\}$. In this case,  the  leading  dipolar interactions~\cite{Flannery_2005,Weber_2017} encompass   pairwise excitation-transfer $(r_i,r_j)=(r,r')\leftrightarrow(r',r)$ even for a vanishing  electric field~\cite{PhysRevA.88.043436,PhysRevLett.110.103001,PhysRevLett.114.113002,PhysRevLett.120.063601,PhysRevLett.119.053202,chen2023spectroscopy}.   This energy transfer can be  described by a dipolar version ${H_{\rm XY}^{\rm dip}}$ of the  XY model~\eqref{eq:model}, with  $J\mapsto J_{ij}$. The couplings $J_{ij}$ display an angular dependence $J_{ij} = J_3(3\cos^2\theta_{ij} -1)/R_{ij}^3$, where $\theta_{ij}$ is the angle between the inter-atomic vector $\boldsymbol{R}_{ij}$ and an external magnetic field $\boldsymbol{B}_0$ (fixing the atoms quantization axis)~\cite{frasier2001resonant,PhysRevLett.114.113002,PhysRevLett.120.063601,doi:10.1126/science.aav9105}.  For instance,  $|J_{ij}|/2\pi\sim 1$-10$\,$MHz for  $^{87}$Rb atoms with $n,n'\sim 50$-60 at $R_{ij}\sim10$-15$\,\mu$m distances, which  exceed the  linewidths set by the Rydberg lifetimes even at room temperatures $T_1\sim100$-$200\,\mu$s~\cite{PhysRevA.72.022347}. 

In principle, the long-range terms could lead to additional gauge-violating processes. Therefore, we selectively address even-even/odd-odd and odd-even interactions. We can suppress the even-even/odd-odd processes by taking the magnetic field $\boldsymbol{B}_0$ along a specific  direction relative to the Rydberg array. For $\boldsymbol{B}_0\!\parallel\!\boldsymbol{R}_{ij}$ ($\boldsymbol{B}_0\!\perp\!\boldsymbol{R}_{ij}$), the interactions are antiferromagnetic $J_{ij}>0$~\cite{PhysRevLett.119.053202,PhysRevLett.114.113002} (ferromagnetic $J_{ij}<0$~\cite{chen2023spectroscopy}). As exploited  in~\cite{doi:10.1126/science.aav9105}, when  $\boldsymbol{B}_0$ is directed along the plane containing a  Rydberg ladder, the XY couplings between distant spins making an angle  $\theta_{ij}=\theta_m:=54.7^{\rm o}$ vanish, yielding approximately a dimerised XY model. 

In our context, this arrangement would suppress the gauge-violating processes involving even-even (odd-odd) spins. On the other hand, the dimerisation would activate  higher-order terms in the high-frequency Floquet expansion,  limiting the timescale of validity of  the QS (see Supplemental Material). To overcome this problem, we 
propose to lift  the quantising field such that $\boldsymbol{B}_0$ is out of plane at an angle $\theta_m$ with the normal vector to the Rydberg ladder (here corresponding to the $\boldsymbol{z}$ axis) - see Fig.~\ref{fig:lattice}. With this choice, besides suppressing the undesired even-even (odd-odd) couplings, the higher-order Floquet terms due to the dimerisation do not contribute. Defining 
 $\phi_{ij}$ as the angle that $\boldsymbol{R}_{ij}$ makes with the  projection of the quantising field into the Rydberg array $\boldsymbol{B}_0^{\perp}$, we find that  $J_{ij} = J_3(2 \cos^2 \phi_{ij}-1)$. Therefore, when $\phi_{ij}=\phi_m:= 45^{\rm o}$, the atoms on the upper (bottom) leg of the ladder  correspond to the vertices of isosceles right triangles, whose sides have length $d$ and whose base have length $b = \sqrt{2} \, d$, which lies along the legs of the ladder. This choice results in $J_{ij} = 0$ for pairs of atoms within the same leg, which correspond to the undesired even-even (odd-odd) spin couplings. The  expression for  the non-zero  XY couplings is 
\begin{equation}
    \label{eq:amplitudes}
    J_{ij} =  -\frac{J_3 \, r}{d^3} \left(\frac{2}{1+r^2}\right)^{\!\!\!5/2}\!\!\!,\hspace{2ex} r = i-j,
\end{equation}
 where the indices $i$ and $j$ must be odd and even, respectively (see Supplemental Material). We thus see that $|J_{i,i\pm 1}|=J_3/d^3$, such that there is no dimerisation, and no higher-order gauge-breaking terms in the Floquet scheme associated to it. Additionally, the couplings $|i-j| > 1$ are highly-suppressed by the dipolar law.

Following the same driving protocol as for the nearest-neighbor Floquet scheme, we obtain the effective Hamiltonian
\begin{equation}
\label{eq:ryd_eff_h}
    H_{\rm eff}=\sum_{i,j,k=1}^{N_a}
\!\frac{J_{ij} J_{jk}}{\omega_{\rm d}}\chi_{ik}\sigma_i^+\sigma_{j}^z\sigma_k^-,
\end{equation}
 where the long-range nature of the XY couplings sets an assisted tunneling involving generic spin triplets (not necessarily nearest-neighbors as in the scheme adopted for ~\eqref{eq:nn_eff_h}). We note that the dressing parameter  $\chi_{ik}$ is still given by Eq.~\eqref{eq:chi}, and we can again exploit the destructive-interference selectivity, such that the tunneling only occurs between odd spins $i\neq k$ mediated by an even spin at $j$. However, there are  additional  longer-range   terms of this form that would break the gauge symmetry, the largest being $\sigma_{2n-1}^+\sigma^z_{2n}\sigma^-_{2n+3}$ and $\sigma_{2n-1}^+\sigma^z_{2n+2}\sigma^-_{2n+1}$. We get rid of the former by appropriately choosing the driving phases according to $\varphi_{2n+1} = n \frac{\pi}{2}$ (see Eq.~\eqref{eq:amplitudes}), while the latter  are much smaller $J_{1,4}/J_{1,2} \approx 0.05$, which follows from Eq.~\eqref{eq:amplitudes}. By sticking to the previous choice for the driving amplitude $\eta=\nu_1$, we find that $\chi_{2n-1,2n+1} = \chi \approx - 0.5  \rm i$. One can  use the effective tunneling in Eq.~\eqref{eq:micro_param}, and map the Floquet-Rydberg QS to the target $\mathbb{Z}_2$ LGT as before.

\begin{figure}
    \centering
    \includegraphics[width = \linewidth]{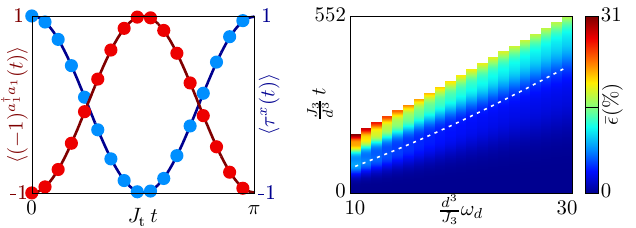}
    \caption{\textbf{Gauge-invariant dynamics and gauge violation.} The left panel displays the driven dynamics (solid line) and the effective gauge-invariant dynamics (dots) arising from the initial state $\ket{L} = \ket{1,-,0}$, in the $N_a=3, \omega_{\rm d} = 30 J_3/d^3$ case. The driven evolution is in full-agreement with the expected periodic oscillations of $\langle a^\dagger_n a_n (t) \rangle$, $\langle \tau^x (t) \rangle$. The right panel corresponds to the average percent error $\bar{\epsilon}(t,\omega_{\rm d})$ accumulated during the evolution from the state $\ket{\psi_0}=\ket{1_1,-,0,- \dots -,0}$, in the case $N_a=9$. For each value of the driving frequency $\omega_d$, only times up to $3\pi/J_{\rm t}(\omega_{\rm d})$ are explored. The dashed line corresponds to a $10 \%$ threshold of gauge violation.}
    \label{fig:single}
\end{figure}

As done for the n.n. model, now we benchmark our scheme by comparing the Floquet effective dynamics with the time-evolution of the original Rydberg system. Due to the long-range nature of the microscopic model, we resort to exact numerical methods,  limiting the system sizes up to $N_a=9$ spins. We consider the initial state with a single charge $\ket{\psi_0} =  \ket{1_1,-,0,-\cdots-,0}$, and calculate numerically the time-evolution operator by a  Trotter expansion within one period of the drive  $U_{T} = \prod_i {\rm e}^{- {\rm i} {H_{\rm XY}^{\rm dip}}(t_i) \delta t}$, where we have taken $\delta t = T/410$ to minimize the numerical errors. The stroboscopic dynamics is obtained as $\ket{\psi(nT)} = U_T^n \ket{\psi_0}$, with $n \in \mathbb{N}$. The results are compared with the effective gauge-invariant dynamics of Eq.~\eqref{eq:z2model}, which have an exact solution in terms of a Wannier-Stark ladder in the single-charge sector. In the case of two matter sites and a single link, we find that the large-frequency $\omega_{\rm d}\gg J$ regime perfectly realizes the expected gauge-invariant evolution (see Fig.\ref{fig:single}). Indeed, the dynamics is restricted to the subspace spanned by $\ket{\rm L} = \ket{1,-,0}$ and $\ket{\rm R} = \ket{0,+,1}$, which are gauge-invariant according to  $G_n = (-1)^{a^\dagger_n a_n} \, \tau^x$, with local charges $q_n=(-1)^n$. The tunneling dynamics leads to Rabi oscillations between $\ket{\rm L}/\ket{\rm R}$, provided that the electric field  is periodically switched on/off $\ket{+}/\ket{-}$ to comply with gauge invariance. As figures of merit, we display the real-time evolution of the $\mathbb{Z}_2$ charge  $\langle (-1)^{a^\dagger_n a_n} \rangle$, and the electric field $\langle \tau^x \rangle$. 

The agreement with the $\mathbb{Z}_2$ gauge-invariant dynamics can be affected by two main sources of error. The first corresponds to gauge-breaking processes that arise from the higher orders in the large-frequency Floquet expansion. The second derives from the long-range XY couplings. To estimate the amount of gauge violation, we extend our simulations to  the case of $9$ spins, and allow to sweep the driving frequency. We evaluate the average percent error $\bar{\epsilon}(t,\omega_{\rm d}) = \frac{100}{N_a} \sum_n \abs{\frac{\langle G_n(t,\omega_{\rm d}) \rangle - q_n}{q_n}}$. As shown in Fig.~\ref{fig:single}, the amount of gauge violation is below a $10 \%$ threshold for values of $\frac{J_3}{d^3} \, t$ that are compatible with currently achieved experimental times. We note that, as $\omega_{\rm d}$ is increased, the agreement between driven and effective dynamics grows, as a result of the $\left( 1/{\omega_{\rm d}}\right)^\ell$ suppression of $(\ell+1)$-th order terms in the high-frequency expansion.

So far, we have focused on the gauge-invariant tunneling stemming from $H_{\rm eff}$~\eqref{eq:ryd_eff_h}. The required longitudinal $\sum_{i \text{ even}}\! \frac{\Omega}{2}\sigma_i^x$ and transverse $\sum_{i \text{ odd}} (-1)^i\frac{\delta h}{2}\hspace{0.2ex}\sigma_{i}^z$ fields, as well as the time-periodic modulation,  require local addressing in the Rydberg platform. This can be achieved by using laser beams suitably diffracted using acousto-optic deflectors. The lasers addressed to the even  atoms should drive Raman transitions with a two-photon Rabi frequency $\Omega$ that is resonant with the two Rydberg energy levels $\omega_{\rm L,1}-\omega_{\rm L,2}=\omega_0$, e.g.~\cite{PhysRevLett.123.230501}. For the odd atoms, on the other hand, a pair of counter-propagating laser beams far off resonant ${\omega}_{\rm L,1}-{\omega}_{\rm L,2}=:\omega_{\rm d}\ll \omega_0$ lead to  differential ac-Stark shifts with contributions stemming from each of the laser beams, which have been previously used to induce a local constant longitudinal field in a two-atom XY model~\cite{PhysRevLett.119.053202}. For our purposes, these local ac-Stark shifts must be tuned  to account for the staggering  $\delta h$. Since we consider a pair of laser beams, there is also a travelling standing-wave pattern coming from their interference, which can be exploited for the time-periodic  longitudinal field. The amplitude  of this modulation $\eta\omega_{\rm d}$ is controlled by the crossed-beam two-photon Rabi frequencies, whereas the phase depends on the wavevectors 
$\varphi_i=(\boldsymbol{k}_{\rm L,1}-\boldsymbol{k}_{\rm L,2})\cdot \boldsymbol{R}_i$. These laser-induced terms can be of the same order as the XY couplings, which requires using high-power lasers to allow for large detunings from other excited states, such that  $\delta h,\Omega,\eta\omega_{\rm d}\sim 0.1$-10$\,$MHz.

{\it Conclusions and outlook}.-- We have devised a protocol for the analog quantum simulation of a $\mathbb{Z}_2$LGT through Floquet engineering, and shown that it can be realised with Rydberg atoms trapped in optical tweezers. The proposed protocol generates specific gauge-invariant 3-body terms~\eqref{eq:z2model} from 2-body gauge-breaking couplings~\eqref{eq:model} by exploiting a selective dressing that makes use of time-periodic Floquet modulation and a destructive interference that depends on the relative inhomogeneous phase of the modulation. The strong dipole-dipole interactions  in Rydberg atoms make them ideally-suited to realise this gauge-invariant dressing, as the weaker second-order processes in Eq.~\eqref{eq:micro_param} can still  be faster than typical decay times. On the other hand, we note that our scheme can be directly  translated to trapped-ion or superconducting-qubit platforms, in which effective XY models have also been demonstrated. Provided that leading noise sources are minimised further, our proposal for a $\mathbb{Z}_2$LGT QS could also work there. 

{\it Acknowledgements}. We thank P. Castorina, P. Kitson, G. Marchegiani, P. Naldesi, F. Perciavalle and J. Polo for discussions. A. B. acknowledges support from PID2021-127726NB-I00 (MCIU/AEI/FEDER, UE), from the Grant IFT Centro de Excelencia Severo Ochoa CEX2020-001007-S, funded by MCIN/AEI/10.13039/501100011033,  from the CSIC Research Platform on Quantum Technologies PTI-001,  from the MINECO through the QUANTUM ENIA project call - QUANTUM SPAIN project,
and from the EU through the RTRP-NextGenerationEU within the framework of the Digital Spain 2025 Agenda.

\bibliographystyle{apsrev4-1}
\bibliography{bibliography}

\onecolumngrid

\newpage
 
\setcounter{equation}{0}


\pagebreak

\widetext

\begin{center}

\textbf{\large Supplemental Material for "A Floquet-Rydberg quantum simulator for confinement in $\mathbb{Z}_2$ gauge theories}

\end{center}



\setcounter{equation}{0}

\setcounter{figure}{0}

\setcounter{table}{0}

\setcounter{page}{1}

\makeatletter

\renewcommand{\theequation}{S\arabic{equation}}

\renewcommand{\thefigure}{S\arabic{figure}}

\renewcommand{\bibnumfmt}[1]{[S#1]}

\renewcommand{\citenumfont}[1]{S#1}

\renewcommand{\thesection}{\Alph{section}}

\section{Floquet scheme for a $\mathbb{Z}_2$ lattice gauge theory on a chain}

Here, we present the derivation of the effective dynamics arising in the high-frequency limit of the driven XY Hamiltonian
\begin{equation}
\label{eq:driven}
    {H}_{XY}(t) = \sum_{i}\sum_{j>i} J_{ij} \!\! \left(\sigma_i^+ \sigma_j^- + {\rm H.c.} \right)+\sum_{i \, \text{even}} \frac{\Omega}{2} \sigma_i^x + \sum_{i \, \text{odd}} (-1)^i \frac{\delta h}{2} \sigma_i^z  + \sum_{i \, \text{odd}}\frac{1}{2} \eta\omega_{\rm d} \cos(\omega_{\rm d} t + \varphi_i) \, \sigma_i^z \, .
\end{equation}

\ni The model accounts for long-range XY interactions, 
together with longitudinal and transverse fields. The transverse field is taken to be non-zero only on even sites, whereas the longitudinal field is non-vanishing only on odd sites. In particular, it  is made up of a staggered contribution $\delta h$, together with  a time-periodic modulation of frequency $\omega_{\rm d}$ and amplitude $\eta \omega_{\rm d}$ that has a crucial inhomogeneous phase $\varphi_i$. Another crucial aspect is that the XY couplings can only couple pairs of spins that reside at even and odd sites, but $J_{ij}=0$ when the indexes of the spin pair are both odd or even. The ideal case fulfilling this condition  would be that of a nearest-neighbor XY model, although it will be important to consider possible longer-range couplings in light of its realization with Rydberg atoms, which is discussed in the next section.

In the case of time-periodic Hamiltonians, the Floquet theorem provides us with a basis of solutions to the associated Schrodinger equation, which is expressed by the so-called Floquet states $\ket{\psi_n(t)}$. In analogy to the case of space-periodic systems, which is based on the application  of Bloch's theorem, Floquet states are defined as the eigenstates of the time-evolution operator over a period $T=\frac{2\pi}{\omega_{\rm d}}$, such that  $U(t_0 + T, t_0) \ket{\psi_n(t_0)} = {\rm e}^{-{\rm i} \epsilon_n T} \ket{\psi_n(t_0)}$. Similarly to Bloch states, Floquet states are written as the product of a "plane-wave", which is fixed by the value of the quasi-energies $\epsilon_n$, and a time-periodic part, which is expressed in terms of  the Floquet modes $\ket{u_n(t)} = \ket{u_n(t+T)}$ as follows 
\begin{equation}
    \ket{\psi_n(t)} = {\rm e}^{-{\rm i} \epsilon_n t} \ket{u_n(t)} \, .
\end{equation}

\ni Any state at time $t$ can thus be expanded in terms of the Floquet states $\ket{\psi(t)} = \sum_n c_n \, {\rm e}^{-{\rm i} \epsilon_n t} \ket{u_n(t)}$. If the initial state corresponds to a specific Floquet mode $\bar{n}$, such that $c_n = 0 \, \, \forall n \neq \bar{n}$, it will evolve periodically and remains in the Floquet mode $\ket{u_{\bar{n}}(t)}$. In general, the time evolution can be factorized as the product of the periodic dynamics of the Floquet modes, which is commonly referred to as micromotion, and deviations from a periodic behaviour due to the interference effects produced by the phase factors ${\rm e}^{-{\rm i} \epsilon_n t}$. When looking at the system at long times with respect to the period $T$, however, the effect of the micromotion can be neglected as its contributions become highly off-resonant. Moreover, the dynamics simplifies considerably when   looking at the system stroboscopically, namely at times that are always integer multiples of $T$. The stroboscopic dynamics is described by the time-evolution operator over a period $T$, which can be written as
\begin{equation}
    U(t_0 + T,t_0) = \exp{-{\rm i} H^{\rm F}_{t_0} T},
\end{equation}

\ni where we have  defined the time-independent Floquet Hamiltonian $H^{\rm F}_{t_0}$, which  depends parametrically and periodically on the starting time of the evolution $t_0$ - see e.g.~\cite{eckardt2015high}. 

In the limit of large driving frequencies  with respect to the other  energy scales involved in the problem ($\hbar=1$), one can evaluate the Floquet Hamiltonian as a power series in the inverse driving frequency.
Since we are interested in a regime of strong driving $\eta >1$, the amplitude of the driving term is larger than its frequency $\omega_{\rm d}$. Therefore, in order to make the perturbative expansion well-defined, we first change into an interaction picture with respect to the time-periodic modulation $\tilde{H}_{\rm XY}^{\rm dip}(t) = U^\dagger(t) {H}_{\rm XY}^{\rm dip}(t) U(t) - {\rm i} U^\dagger(t) \partial_t U(t) $, where $U(t) = {\rm exp}\{-i \sum_{j} \frac{\Gamma_j(t)}{2} \, \sigma_j^z\}$ and $\Gamma_j(t) = \eta \sin(\omega_{\rm d} t + \varphi_j)\delta_{j,{\rm odd}}$, where we have neglected an irrelevant time-independent phase that can be actually be gauged away by a trivial $U(1)$ transformation on the ladder operators. The resulting Hamiltonian reads
\begin{equation}
    \label{eq:rotated_model}
    \tilde{H}_{\rm XY}^{\rm dip}(t) =  \sum_{i}\sum_{j>i}J_{ij} \left({\rm e}^{{\rm i} \eta \sin(\omega_{\rm d} t + \varphi_i)} \sigma_i^+ \sigma_j^- + {\rm H.c.}\right)+\sum_{i \, \text{even}} \frac{\Omega}{2} \, \sigma_i^x + \sum_{i \, \text{odd}} (-1)^i \frac{\delta h}{2} \sigma_i^z, 
\end{equation}

\ni which now manifestly depends only on small couplings under the assumption $\Omega, J_{ij} \ll \omega_{\rm d}$. Denoting $\tilde{H}_\ell = (1/T) \int_0^T {\rm e}^{-{\rm i} \ell \omega_{\rm d} t} \, \tilde{H}_{\rm XY}^{\rm dip}(t)$, the stroboscopic dynamics, up to second order in the large-frequency expansion, is described by the effective Hamiltonian $\tilde{H}_{\text{eff}} = \tilde{H}_0 + \sum_{\ell > 0} \frac{1}{\ell \omega} \commutator{\tilde{H}_\ell}{\tilde{H}'_{-\ell}}$, which is unitarily related to the Floquet Hamiltonian ${H}_{t_0}^F$ introduced above - see \cite{eckardt2015high}. Making use of the Jacobi-Anger expansion~\cite{bessel_book}, and computing the corresponding commutators, we find that
\begin{equation}
    \nonumber
    \tilde{H}_{\text{eff}} \approx \sum_{i \text{ even}} \frac{\Omega}{2} \, \sigma_i^x + \sum_{i \, \text{odd}} (-1)^i \frac{\delta h}{2} \sigma_i^z + \sum_{i,j>i} J_{ij}  \mathsf{J}_0(\eta) \big(\sigma_i^+ \sigma_j^- + {\rm H.c.}\big) - 2 \text{i} \sum_{i,j,k} \sum_{\ell > 0} \frac{\mathsf{J}^2_\ell(\eta)}{\ell} \sin[\ell(\varphi_i -\varphi_k)] \frac{J_{ij} J_{jk}}{\omega_{\rm d}} \sigma_i^+ \sigma^z_j \sigma_k^- .
\end{equation}
 Note that first-order corrections amount to a renormalization of the hopping strength by the  $\mathsf{J}_0(\eta)$ Bessel function. At second order, we obtain assisted tunneling processes in which spin excitations hop between two sites, assisted by the value of the local magnetization at some other site. Recalling that $J_{ij}\neq 0$ only for indexes of different parity, we have processes where a spin tunnels between a pair of odd sites and gets mediated by a spin at an even site, or viceversa. We note that this tunnelings can bring the spin excitation back to the same site, amounting to a 2-body spin-spin coupling of the Ising type. 

Since all the different contributions with $\ell>0$ vanish for $\varphi_i-\varphi_k=0$, one immediately sees that the assisted tunneling can only take place between a pair of different of sites, and be mediated by an intermediate spin at an even site. This is the destructive intereference aluded to in the main text, which allows us to achieve the desired selectivity on the dressing. In addition, if we choose $\eta = \nu_1$, where $\nu_1$ is the first zero of $\mathsf{J}_0(\eta)$ and $\varphi_{2j+1} = j \, \varphi$, we obtain the simpler Hamiltonian
\begin{equation}
    \label{eq:finaleff}
    \tilde{H}_{\text{eff}} \approx \sum_{i \text{ even}} \frac{\Omega}{2} \, \sigma_i^x + \sum_{i \, \text{odd}} (-1)^i \frac{\delta h}{2} \sigma_i^z + \sum_{i,k \text{ odd}}\sum_{ j \text{ even}}  \, \frac{J_{ij} J_{jk}}{\omega_{\rm d}}\chi_{ik} \sigma_i^+ \sigma_j^z \sigma_k^-,
\end{equation}
where we find the dressing parameter defined in the main text $\chi_{ik} = \sum_{\ell>0} \frac{2 \text{i}}{\ell} \mathsf{J}^2_\ell(\eta) \sin(\ell(k-i) \, \varphi/2)$. 

At this point, we can particularise to the ideal case of nearest-neighbor interactions. by identifying odd (even) spin variables with matter (gauge) degrees of freedom
of a halved chain, one can define  harcore bosonic matter and $\mathbb{Z}_2$ gauge fields as  follows $\displaystyle \sigma_{2i+1}^+=a^\dagger_i,\,\,\,\hspace{1ex}\sigma_{2i+1}^-=a^{\phantom{\dagger}}_i,\,\,\,\hspace{1ex}\sigma_{2i+1}^z=2a^\dagger_ia^{\phantom{\dagger}}_i-1,
\sigma_{2i}^z=\tau^z_{i+1/2},\hspace{1ex} \sigma_{2i}^x=\tau^x_{i+1/2}$, and taking $J_{\text{t}} = {J^2} \chi/\omega_{\rm d}$, $h = {\Omega}/{2}$, $\mu = {\delta h}$, we obtain the desired model of the $\mathbb{Z}_2 LGT$  in Eq.~(1) of the main text. Note that, for nearest neighbor interactions, $\chi$ does not bear any further dependence on the site indices, and we fix it by choosing the value of $\varphi = \pi/3$, which maximizes the renormalized coupling $J_\text{t}$ - see Fig.(2) in the main text. For Rydberg-atom platforms, where the couplings have a dipolar decay with the distance, there are additional consideration that need to be addressed, as discussed in the following section.

\section{Rydberg-atom scheme for the $\mathbb{Z}_2$LGT quantum simulator}

In this section, we discuss the several key points for the experimental realisation of the Floquet scheme.
 Since  the selective dressing lowers the effective interactions by, at least, the factor  $J/\omega_{\rm d}$,  the targeted dynamics might get too slow in comparison to leading sources of decoherence in various experimental platforms that can realise spin-spin model. From this perspective, Rydberg platforms are very attractive, as the spin-spin couplings are much larger than the leading noise sources. In this part of the SM, we provide additional details that support the arguments of the main text. We provide below a detailed account on current Rydberg-atom QSs, which serves to set the notation and present important tools that are used in the main text.
 
 The spin-1/2 system  can be implemented using Rydberg qubits, which  are encoded in the level structure of neutral atoms, such as the heavy alkali   $^{87}$Rb and $^{133}$Cs, or the alkaline-earth $^{88}$Sr. The spin/qubit subspace $\ket{0}=\ket{\uparrow},\ket{1}=\ket{\downarrow}$ is composed of groundstate hyperfine level(s) $\{\ket{g}\}$ and, additionally, contains or admixes  highly-excited  Rydberg state(s) $\{\ket{r}\}$~\cite{Browaeys_2016,10.1116/5.0036562}. Due to the large principal quantum number of the later  $n\sim40\,$-$\,80$, there are very strong Rydberg-Rydberg interactions $V_{\rm dd}$  caused by  the  large dipole matrix elements $\bra{r}d\ket{r'}\propto n^2ea_0$, where $e$ is the valence electron charge and $a_0$ the Bohr radius. In addition, these large values of $n$ also underlie  the typical  long decay times, e.g. $T_1\sim200\,\mu$s for $^{87}$Rb and $n=80$~\cite{PhysRevA.72.022347}. These two properties contribute to a characteristic regime   $V_{\rm dd}\gg T_1^{-1}$, which lies at the very heart of  seminal proposals for  entangling gates~\cite{PhysRevLett.85.2208,PhysRevLett.87.037901} that have  stimulated the  experimental progress~\cite{PhysRevLett.104.010502,PhysRevLett.104.010503} towards the current high-fidelity Rydberg-atom processors~\cite{PhysRevLett.123.170503,Madjarov2020}.
 
 The capability of creating  qubit arrays  by loading one atom per site    makes this platform   an interesting     QS, which can be achieved by either working  in the Mott-insulating regime of an optical lattice~\cite{Schauß2012,doi:10.1126/science.1258351,Zeiher2016}, or by assembling the atoms one by one  in reconfigurable  tweezer arrays~\cite{doi:10.1126/science.aah3752,doi:10.1126/science.aah3778}. 
When the inter-atomic distances are much larger than the radius of the Rydberg states,  the  overlap of the weakly-bound valence electrons becomes negligible, and the  interactions are accurately described  by the electrostatic potential between the corresponding  charge distributions. Since  the Rydberg atoms are neutral, a multipole expansion is  dominated by the dipole-dipole interactions~\cite{Flannery_2005,Weber_2017}, which   can induce  transitions between  pairs of    Rydberg levels $r_i^{\phantom{'}},r'_j\to r_i'',r_j'''$, where $i,j\in\{1,\cdots N_a\}$ label the pair of atoms involved.  These transitions   can be exploited to simulate various spin models. For instance, in the   $gr$ scheme, the qubits are encoded in a single Rydberg (groundstate) level  $\ket{0_i}=\ket{r_i}$ ($\ket{1_i}=\ket{g_i}$) for each of the atoms,    yielding a typical transition frequency  $\omega_0/2\pi\sim$1000$\,$THz for states with vanishing angular momentum $L=0$. Off-resonant transitions from the Rydberg states to neighboring ones   yield, at second order,   a Van der Waals energy shift corresponding to an  Ising-type interaction ${H}_{\text{Ising}} = \sum_{i,j} V_{ij} \ (1+\sigma_i^z)(1+ \sigma_j^z)$. Here, $V_{ij}= V_6/{R_{ij}^6}$  decays with  the  distance  between the ionic cores of a pair of atoms $R_{ij}$, and the  parameter $V_6$ depends on the specific atomic species and Rydberg levels. For instance, for $^{87}$Rb atoms with $L=0$,  $V_6/2\pi\approx 4.6 n^{11}\,$nHz$\cdot\mu$m$^6$~\cite{Browaeys_2016}. The  extreme scaling with the principal quantum number, which  results from both the large transition dipoles and the small energy difference between nearby Rydberg levels $n\pm 1$, $\ell=1$~\cite{PhysRevA.72.022347}, is responsible for the strong Ising interactions e.g. $V_{ij}/2\pi\approx 4\,$MHz$\,\gg T_1^{-1}$ for $n=80$  at $10\mu m$~\cite{Browaeys_2016,PhysRevA.72.022347}. In addition to these interactions, 
a laser-induced Rabi driving of frequency $\omega_{\rm d}$ leads, in a rotating frame, to the term  ${H}_{\text{field}}=\frac{1}{2}\sum_i(\Omega\sigma^x_i+\Delta\sigma_i^z)$, where  $\Omega$ is the Rabi frequency, and   $\Delta=\omega_0-\omega_{\rm d}$ is the detuning. Altogether, ${H}_{\text{Ising}}+{H}_{\text{field}}$ yields a QS of a   long-range quantum Ising model in both transverse $h_t=\Omega/2$ and longitudinal $h_l=\Delta/2$ fields. This Ising-model QS has been realised  both in optical lattices~\cite{Schauß2012,doi:10.1126/science.1258351} and tweezer arrays~\cite{Labuhn2016,Bernien2017}. We note that, as  $R_{ij}$ is lowered, one  gets closer to  other Rydberg levels, and the distance dependence of $V_{ij}$ changes towards a  dipolar decay~\cite{RevModPhys.82.2313}.

Let us now discuss a different possibility to engineer  spin models, which is based on resonances through the dipole-dipole interactions. Some experiments focus on the so-called  F\"{o}rster resonances~\cite{PhysRevLett.97.083003,PhysRevLett.100.233201,doi:10.1126/science.1244843,Ravets2014}, which   involve transitions between several nearby Rydberg levels, which  can be tuned on resonance via the stark shift under a weak  electric field~\cite{PhysRevA.72.022347}.  In this work, we shall instead focus on the
$rr$ scheme, which encodes the qubit in only a pair of nearby  Rydberg levels $\ket{0}_i=\ket{r}_i$,  $\ket{1}_i=\ket{r'}_i$ with opposite  parity, e.g. $ L'=L\pm1$. In this case, the typical qubit frequencies are $\omega_0/2\pi\sim 10\,$GHz~\cite{Browaeys_2016}, and one can induce the Rabi drivings  using microwaves or laser-induced  Raman transitions~\cite{10.1116/5.0036562}. The dipole-dipole interactions  lead naturally  to a  resonant  channel for energy transfer  $(r_i,r_j)=(r,r')\to (r', r)$ and, thus, dispense with the need of a Stark-tuned resonance~\cite{PhysRevA.88.043436,PhysRevLett.110.103001,PhysRevLett.114.113002,PhysRevLett.120.063601,PhysRevLett.119.053202}.   The   interactions can be described by a long-range isotropic XY model ${H}_{\text{XY}}^{\rm dip} = \sum_{i,j>i} J_{ij} (\sigma^+_i \sigma^-_j + {\rm H.c.})$ where, in contrast to the previous case,    $J_{ij}= {J_3}/{R_{ij}^3}$  arises at first order and thus decays with a dipolar law.  For $^{87}$Rb with $n=63, n'=62$,  $J_3/2\pi\approx 0.5n^4\,$kHz$\cdot\mu$m$^3$~\cite{PhysRevLett.114.113002}, which would  lead to  $J_{ij}/2\pi\approx 7$MHz$\,\gg T_1^{-1}$ at $10\,\mu$m distances.

The remaining ingredients to arrive at a dipolar version of Eq.~\eqref{eq:driven} require local addressing in the drivings of even and odd qubits to realise  the specific longitudinal and trasverse fields. As discussed in the main text, in Rydberg platforms, this  can be achieved by using addressed laser beams that induce resonant Raman drivings of the even Rydberg atoms, and a combination of static and time-periodic off-resonant ac-Stark shifts. We shall be interested in a regime $\delta h,\Omega,\eta\omega_{\rm d}\sim 0.1$-10$\,$MHz.
In order to apply the Floquet scheme described in the previous section, we first need to remove all possible processes that couple odd to odd and even to even spins. To do so, we consider a ladder configuration depicted in Fig. 1 of the main text, and set an external magnetic field $\boldsymbol{B}_0$ which defines the quantization axis, in such a way to make the spin-spin couplings anisotropic: $J_{ij} = {J_3} (3\cos^2\theta_{ij}-1)/{R_{ij}^3}$,  $\theta_{ij}$ being  the angle between the inter-atomic vector $\boldsymbol{R}_{ij}$ and  $\boldsymbol{B}_0$~\cite{frasier2001resonant,PhysRevLett.114.113002,PhysRevLett.120.063601,doi:10.1126/science.aav9105}. We note that other angular dependencies can be found for different F\"{o}rster resonances involving more levels in Rydberg platforms~\cite{PhysRevLett.93.153001,PhysRevA.92.020701,nipper_thesis}.
Identifying the atoms lying on the bottom (upper) leg of the ladder with odd (even) numbers - see Fig.(1) in the main text, we can align the legs of the ladder along a critical direction for which $\theta_{ij} = \theta_m = \arccos{1/\sqrt{3}}$ when $i,j$ both odd (even), such that we  suppress all odd-odd (even-even) couplings~\cite{doi:10.1126/science.aav9105}. 

Resorting to the scheme described in the main text, we can describe the system of Rydberg atoms according to the Hamiltonian \eqref{eq:driven}. Proceeding with the high-frequency expansion, as described in the previous section, we arrive at Eq.\eqref{eq:finaleff}. To achieve the identification with the $\mathbb{Z}_2$ LGT, we need to control  longer-range couplings and make them small enough, as they contribute with gauge-breaking terms. We achieve this purpose by suitably adjusting the geometry of the system and the lifting the external magnetic field $\boldsymbol{B}_0$ out of the plane of the atoms, going beyond previous situations ~\cite{doi:10.1126/science.aav9105}.
We take $\boldsymbol{B}_0$ to form an angle $\theta_m$ with the $\boldsymbol{z}$ axis, where the atoms lie on the $x$-$y$ plane. We choose the ladder to be parallel to the $\boldsymbol{x}$ direction. Identifying as $\boldsymbol{B}_0^\perp$ the perpendicular projection of $\boldsymbol{B}_0$ on the $x$-$y$ plane, we can rewrite the XY couplings in terms of the angle $\phi_{ij}$ that $\boldsymbol{B}_0^\perp$ forms with the inter-atomic vector $\boldsymbol{R}_{ij}$ using the following identities \begin{equation}
    \cos(\theta_{ij}) = \frac{\boldsymbol{B}_0 \cdot \boldsymbol{R}_{ij}}{B_0 \, R_{ij}} = \frac{\boldsymbol{B}_0^\perp \cdot \boldsymbol{R}_{ij}}{B_0 \, R_{ij}} = \frac{B_0^\perp}{B_0} \cos(\phi_{ij}) = \sin(\theta_m) \cos(\phi_{ij}) = \sqrt{\frac{2}{3}} \cos(\phi_{ij}).
    \end{equation}
     As a result, the spin-spin couplings have the following angular dependence
\begin{equation}
    J^{ij}_{\text{dd}} = \frac{J_3}{R_{ij}^3} (2 \cos^2\phi_{ij}-1) \, .
\end{equation}

\ni Critical directions for which $\theta_{ij} = \arccos{1/\sqrt{3}}$ now correspond to $\phi_{ij} = \pm \pi/4, \pm 3\pi/4$. Our requirement that the odd (even) atoms lie along critical directions then amounts to aligning the external magnetic field in such a way that $\boldsymbol{B}_0^\perp$ makes an angle of $\pi/4$ with the $\boldsymbol{x}$ axis - see Fig.(1) of the main text. We fix the geometrical arrangement of the atoms by placing them on the vertices of isosceles right triangles with sides of length $\rm d$ parallel to $\boldsymbol{B}_0^\perp$ and basis $b = \sqrt{2} \rm d$ lying on one of the legs of the ladder. 

We can evaluate $J_{ij}$ for any $i,j$. Fixing $i$ to be odd and $j$ to be even, we define $r = i-j$. Hence, we calculate $R_{ij}$ and $\cos(\phi_{ij})$ for any $i,j$. The length of the inter-atomic distance $R_{ij}$ is given by the the hypotenuse of right triangles of basis $\ell_r= \abs{r} \frac{b}{2} = \abs{r} \frac{\rm d}{\sqrt{2}}$ and height $\rm h = \frac{d}{\sqrt{2}}$, which is given by $R_r = \text{d}\left(\frac{1+r^2}{2}\right)^{1/2}$. Defining the angle that $R_r$ forms with the $\boldsymbol{x}$ axis as $\alpha_r$, we see that $\phi_r = \frac{\pi}{4} - \alpha_r$ for $r<0$, or $\phi_r =\frac{3\pi}{4} - \alpha_r$ for $r>0$. Owing to $\cos(\alpha_r) = \frac{\ell_r}{R_r} = \frac{\abs{r}}{\sqrt{1+r^2}}$, we find $\cos{\phi_r} = \frac{1}{\sqrt{2}} \frac{1 - r}{\sqrt{1+r^2}}$, and   finally get to the desired expression for the dipolar XY couplings between even and odd spins
\begin{equation}
    J_{i,i+r} = - \frac{J_3 \, r}{d^3} \left( \frac{2}{1+r^2} \right)^{5/2} \, .
\end{equation}

\ni The last result shows that the couplings are anti-symmetric with $r$, and they decay fast enough for us to neglect processes beyond nearest-neighbors, within reasonable time-scales. Indeed, next to nearest neighbor processes are characterized by an amplitude $J_{1,4} = \frac{3}{5^{5/2}} J_{1,2} \approx \frac{1}{18.6} J_{1,2}$.
In order to further suppress unwanted processes in the effective dynamics, we have made the choice $\varphi = \frac{\pi}{2}$, which fixes the values of $\chi_{ik}$ in Eq.\eqref{eq:finaleff} to zero for $i-k = 4 n$, with $n \in \mathbb{Z}$.
As a result, the effective model obtained from the Floquet scheme will correspond to the target $\mathbb{Z}_2$ theory, with tunneling parameter $J_{\rm t} = \frac{(J_{1,2})^2}{\omega_d} \chi$, where $\chi = \chi_{1,3}$, except for corrections smaller than or equal to $\approx \frac{J_{\rm t}}{18.6}$. 

\section{Floquet errors and gauge violation in the quantum simulator}
The numerical results that we have presented in the main text show the dynamics obtained from the full driven Hamiltonian  in Eq.~\eqref{eq:driven}. The agreement with the gauge-invariant dynamics provided by the target theory (Eq.(1) of the main text) can be benchmarked by calculating the expectation values of the $\mathbb{Z}_2$ gauge-symmetry generators $G_n$. We have estimated the agreement to the target theory through the average percent error $\bar{\epsilon}(t,w_{\rm d}) = \frac{100}{N} \sum_n \abs{\frac{\langle G_n(t,\omega_{\rm d}) \rangle - q_n}{q_n}}$, where $q_n$ are the eigenvalues of the local generators $G_n$ according to the $\mathbb{Z}_2$ LGT, which are fixed by the choice of the initial state of the evolution $\ket{\psi_0}$. Possible sources of gauge-violation arise from either higher-order contributions to the large frequency expansion, or from the effects of the long-range couplings, which can become relevant at sufficiently-long  times. The former can be suppressed by employing larger values of the driving frequency $\omega_{\rm d}$ - see Fig.(4) of the main text. The first neglected contributions come from third-order terms in the Floquet expansion: they either result from three successive elementary tunnelings $\mathcal{O}(J_{ij} J_{jk} J_{kl}/\omega_{\rm d}^2)$ or from processes that would scale with $\mathcal{O}(J_{ij}J_{jk}\Omega/\omega_{\rm d}^2,J_{ij}J_{jk}\delta h/\omega_{\rm d}^2)$, which become negligibly-small in the regime of interest in which $\Omega,\delta h$ and $J_{i,i+1}^2/\omega_{\rm d}$ have the same order of magnitude. Moreover, our choice of the geometry determines a suppression of the third-order terms which grow with the anisotropy between the amplitudes of the elementary contributions to the second-order gauge-invariant tunnelings.
With the devised configuration, the aforementioned elementary processes have indeed the same order of magnitude.  
\begin{figure}[th!]
    \centering
    \includegraphics[width = 0.4\linewidth]{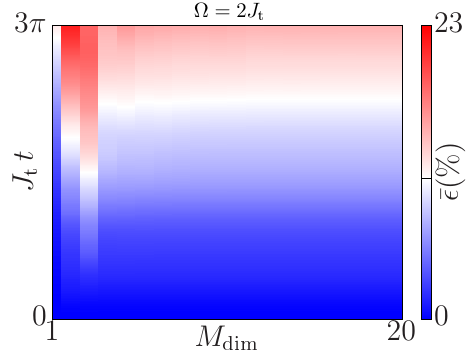}
    \caption{\textit{Plot of the average gauge-violation $\bar{\epsilon}$ as a function of time and of the allowed maximal bond dimension.} The plot refers to the same configuration as for Fig.(2) of the main text. The data is plot at integer multiples of T.}
    \label{fig:err_mps}
\end{figure}
The results obtained by applying the Floquet scheme to a nearest-neighbor XY model are shown in Fig.(2) of the main text, and are obtained using the TEBD algorithm on MPS of fixed maximal bond dimension \cite{itensor,itensor-r0.3,perez2006matrix,banuls2023tensor}. To reduce the numerical error arising from the trotterization, we employ elementary time steps of $\delta = T/400$, $T$ being the period of the drive. The convergence of the method has been tested by showing that the obtained results agree after a certain threshold is reached in increasing the allowed maximal bond dimension - see Fig.\ref{fig:err_mps}.

\end{document}